\documentclass{article}

\PassOptionsToPackage{numbers,compress}{natbib}
\usepackage[preprint]{neurips_2026}
\usepackage[utf8]{inputenc} 
\usepackage[T1]{fontenc}    
\usepackage{hyperref}       
\usepackage{url}            
\usepackage{booktabs}       
\usepackage{amsfonts}       
\usepackage{amsmath}        
\usepackage{nicefrac}       
\usepackage{microtype}      
\usepackage{xcolor}    
\usepackage{graphicx}
\hypersetup{colorlinks=true,linkcolor=blue!60!black,citecolor=blue!60!black,urlcolor=blue!60!black}

\title{Learned Graph Bases for Neural Operators on\\
Power-System Transients}

\author{%
  Shubhan Bhattacharya\\
  University of Pennsylvania\\
  Philadelphia, PA 19104, USA\\
  \texttt{shubhanb@engineering.upenn.edu}\\
  \And
  Avi Bagchi\\
  The University of Texas at Austin\\ Austin, TX 78712, USA\\
  \texttt{aviba@utexas.edu}\\
}

\begin{document}

\maketitle

\begin{abstract}
Fourier neural operators predict power-system transient trajectories
with high accuracy. We extend this line of work by examining the
choice of spatial spectral basis, studying both the graph Fourier
transform derived from the grid's own coupling network and learnable
bases trained from that starting point and from a Fourier control.
Evaluated in one controlled pipeline across three protocols spanning
unseen fault locations and shifted operating regimes, the learnable
bases improve on a parameter- and budget-matched DFT variant by
$7$--$14\%$ in distribution, with per-case significance on both
in-distribution protocols and no loss on the third; the gain comes
from adaptation rather than from the coordinates alone. We further
characterize the regime dependence of this gain, identifying its
cause, locus, and continuity, and show that a resolution requires no
new model: averaging the learned and frozen variants gives the best
shift-robust predictor, while their disagreement provides a
label-free alarm that flags every shifted case with no false alarms.
\end{abstract}

\section{Introduction}
\label{sec:intro}

Reliable grid operation depends on knowing, within seconds of a
disturbance, whether the system will return to synchronism.
Learning-based surrogates promise orders-of-magnitude speedups for
this assessment, and Fourier neural operators (FNOs)
\cite{li2021,kovachki2023} predict post-fault generator trajectories
with striking accuracy \cite{cui2024}. As these models mature, the
question that determines their path toward a control room is not
in-distribution accuracy on a static test split, but behavior under
the conditions a real grid supplies: unseen fault locations, shifted
operating regimes, and changing topology. We study one
degree of freedom, the choice of spatial spectral basis, under that style of evaluation.

Our starting point is a property of the physics. Linearizing the
swing equations yields an inertia-weighted network Laplacian $\Gamma$
whose eigenbasis, the graph Fourier transform (GFT) of the grid
\cite{shuman2013,ortega2018}, is the representation in which the
linearized dynamics decouple. Spectral operators in this domain have
instead used a discrete Fourier transform along the generator axis,
the eigenbasis of a ring graph in machine-index order rather than of
the grid's actual coupling network. We implement the graph
eigenbasis, extend it to a \emph{learnable} orthogonal rotation
initialized at the physics, and characterize what the basis
contributes. Our contributions:

\begin{enumerate}\setlength\itemsep{1pt}
    \item Prior work applies a DFT along the generator axis; we apply
  a GFT instead, using the grid's own Laplacian eigenbasis and a
  learnable rotation initialized there.
    \item At matched parameters and budget across three protocols,
  learnable bases beat a matched DFT by $7$--$14\%$ relative $L^2$
  in distribution; frozen physics bases alone are inconsistent.
  \item Under zero-shot shift the learned basis loses on 41 of 41
  stressed cases; we trace this to training exposure, localize it to
  the basis and its co-adapted weights, and show the trade is a
  continuous dial.
\end{enumerate}
\section{Method}

\subsection{Spectral neural operators and the choice of spatial basis}
\label{subsec:basis}

We build on the Fourier Neural Operator (FNO) formulation of
Cui et al.~\cite{cui2024}, which predicts post-fault trajectories of
generator rotor angle $\delta$, frequency deviation $\omega$, and
voltage $V$ on the NPCC 48-machine system. Each FNO layer transforms
its input to a spectral representation, applies learned per-mode
multipliers to a truncated set of coefficients, and transforms back
(Fig.~\ref{fig:arch}). Training loss itself remains in the time domain. Along the
temporal axis this construction is natural: the discrete Fourier basis
diagonalizes shift-invariant operators, and post-fault dynamics are
dominated by a band of electromechanical oscillations. Along the
\emph{generator} axis, however, the choice of basis carries a hidden
assumption. Applying a DFT across the 48 machines treats the generator
index as a spatial coordinate on a periodic ring, so that
``low-frequency'' modes are smooth in an arbitrary ordering:
machine 17 is not physically adjacent to machines 16 and 18. The
spectral truncation that gives the FNO its
efficiency is therefore performed in coordinates with no relationship
to the network that actually couples the generators.

\subsection{A graph basis from the swing equations}
\label{subsec:graph}

The physics suggests the correct coordinates, as the baseline's own
derivation identifies \cite[Eq.~7]{cui2024}. Linearizing the classical
swing equations \cite{kundur1994,sauer1998} about an
operating point gives
\[
M\ddot{\delta} \;=\; -\,CBC^{\top}\delta \;+\; p,
\]
where $M=\mathrm{diag}(m_i)$ collects machine inertias, $B$ contains
line susceptances, and $C$ is the incidence matrix of the reduced
network, so that $L = CBC^{\top}$ is a weighted graph Laplacian.
Substituting $\delta = M^{-1/2}x$ symmetrizes the system, and the
resulting operator
\begin{equation}
\Gamma \;=\; M^{-1/2}\,C B C^{\top} M^{-1/2},
\qquad
\Gamma = U\Lambda U^{\top},
\label{eq:gamma}
\end{equation}
is diagonalized by an orthonormal eigenbasis $U$ whose columns are the
grid's electromechanical mode shapes: the zero-eigenvalue column is
the inertia-weighted common (center-of-inertia) mode, and increasing
eigenvalues correspond to inter-area swings through increasingly
localized machine oscillations. Replacing the generator-axis DFT with
this eigenbasis, a graph Fourier transform (GFT) with respect to
$\Gamma$, makes the FNO's low-mode truncation retain the modes the
physics designates as dominant, rather than modes smooth in an
arbitrary labeling.

\subsection{A learnable rotation of the physics basis}

Real grids do not behave exactly like the linearized equations of
\S\ref{subsec:graph}: the true coupling between machines is a sine
that only looks linear for small swings, its strength changes with
loading, and voltage is not in the model at all. The eigenbasis $U$ therefore diagonalizes an idealization, not the
real system, which makes it a good starting point but a poor
constraint. We
follow \cite{lezcano2019},
\[
U(\theta) \;=\; \exp\!\big(W - W^{\top}\big)\,U_{\mathrm{init}},
\]
where $W \in \mathbb{R}^{48\times 48}$ is a generator trained jointly
with all other parameters by standard backpropagation, shared across
the spectral layers. Because the matrix exponential of a
skew-symmetric matrix is orthogonal, $U(\theta)$ has orthonormal
columns for \emph{every} value of $W$: the constraint is built into
the parameterization rather than enforced by penalties, and
analysis/synthesis by transpose remains exact throughout training. At
$W=0$ the basis equals $U_{\mathrm{init}}$, so optimization starts at
the physics and every gradient step is a rotation away from it. Each
layer retains the leading $k$ columns of the rotated basis; since the
rotation acts on the full frame \emph{before} truncation, the model
learns not only per-mode multipliers but the pass-band itself. In the
mass-consistent variant, analysis and synthesis carry the
$M^{\pm 1/2}$ weighting of \S\ref{subsec:graph}, preserving the inertia geometry.

Two initializations define the comparison: the graph basis of \S\ref{subsec:graph}
(\emph{gftinit}) and the real Fourier basis (\emph{fourierinit}), a
physics-free control. Parameter counts match to within $0.2\%$
(Table~\ref{tab:models}), and $W$ is excluded from weight decay,
which would bias the basis toward its initialization. Two naming
notes: the frozen graph reference is energy-rotated to diagonalize
the training covariance, so we label it constrained POD rather than a
GFT; and the Fourier initialization's constant first column
reconstructs to $M^{-1/2}\mathbf{1}$ rather than uniform motion, a
common-mode difference we flag wherever initializations are compared.

\subsection{Experimental setup}

The model receives a short post-fault
window of all generator states with fault-type and fault-line
indicators, and predicts $(\delta,\omega,V)$ over a 4.5\,s horizon
($48\times3\times150$: generators, states, timesteps).
Data are simulated three-phase faults on the NPCC system: a
nominal-loading pool and a smaller heavily loaded (\emph{stressed})
pool containing nearly all unstable outcomes. All models share one
operating point (four spectral layers of width 20; 12 generator-axis,
9 temporal, and 3 state-axis modes) and one recipe (AdamW,
warmup--cosine, 800 updates, channel-global normalization, best
checkpoint by validation error), fixed from an earlier screening
phase. Table~\ref{tab:models} lists the five models; a two-model
ensemble is defined in Table~\ref{tab:main}.

\begin{table}[t]
  \caption{Models compared. All share the operating point and recipe
  above; the parameter difference is the shared rotation generator
  $W$. Frozen GFT runs on Protocols A--C were added after the initial
  studies, under the same recipe, seeds, and paired tests.}
  \label{tab:models}
  \centering
  \small
  \begin{tabular}{llllr}
    \toprule
    Model & Generator-axis basis & Trained? & Source of basis & Params \\
    \midrule
    DFT               & complex Fourier            & frozen  & ring assumption (\S\ref{subsec:basis})        & 1{,}040{,}369 \\
    Frozen GFT        & eigenbasis of $\Gamma$     & frozen  & physics (\S\ref{subsec:graph})                & 1{,}040{,}369 \\
    Frozen GFT (POD)  & energy-rotated eigenbasis  & frozen  & physics $+$ training data      & 1{,}040{,}369 \\
    Learned (gftinit) & $U(\theta)$, Eq.~above     & trained & init: physics (\S\ref{subsec:graph})          & 1{,}042{,}673 \\
    Learned (fourierinit) & $U(\theta)$, Eq.~above & trained & init: real Fourier             & 1{,}042{,}673 \\
    \bottomrule
  \end{tabular}
\end{table}

\subsection{Evaluation protocols}

Three protocols differ in exactly one dimension each (200 test cases
each).

\begin{itemize}\setlength\itemsep{2pt}
  \item \textbf{A: unseen fault locations.} Faults on four
  transmission lines never faulted in training, with an entirely
  stable, nominal-loading test set.
  \item \textbf{B: zero-shot regime shift.} The same construction,
  with the split seed chosen by a pre-registered, results-blind rule
  that places stressed cases in test; because they cluster on few
  lines, training then contains \emph{no} stressed cases, making the
  41 stressed test cases (31 unstable) a pure zero-shot shift.
  \item \textbf{C: exposure counterpart.} A uniform random split, so
  test lines appear in training and stressed cases are
  in-distribution; shift-driven failures under B should vanish here.
\end{itemize}

\paragraph{Metrics and statistics.} We report per-case joint relative
$L^2$ error over $(\delta,\omega,V)$. Test composition moves
aggregate error more than any modeling choice studied, so every
aggregate is stratified by case source (nominal/stressed), with
stability labels from a $0.2$\,Hz criterion. Comparisons are paired
per test case (Wilcoxon signed-rank; bootstrap CI on the mean
difference), with three seeds per model and protocol and per-case
errors seed-averaged before pairing. A comparison is significant
only if the Wilcoxon test rejects and the bootstrap CI excludes
zero. Decision rules for Protocol~B
and the mitigation experiments were fixed before results existed.

\section{Results}

Table~\ref{tab:main} reports mean relative $L^2$ error (three
seeds), stratified by case source. Two regimes organize the
results: in distribution, learnable bases help consistently and
frozen bases do not; under zero-shot shift the ordering inverts, and no single
model wins both.

\begin{table}[t]
  \caption{Mean relative $L^2$ error (3 seeds) by protocol and
  stratum (B: 159 nominal / 41 stressed; C: 189 / 11). Bold marks the
  best \emph{single} model per column; the ensemble (per-case average
  of Learned (gftinit) and Frozen GFT (POD)) is shown for reference.
  C's stressed column has $n{=}11$ and is descriptive.}
  \label{tab:main}
  \centering
  \small
  \begin{tabular}{lccccc}
    \toprule
    & \multicolumn{1}{c}{A} & \multicolumn{2}{c}{B} & \multicolumn{2}{c}{C} \\
    \cmidrule(lr){2-2}\cmidrule(lr){3-4}\cmidrule(lr){5-6}
    Model & stable & nominal & stressed & nominal & stressed \\
    \midrule
    DFT                   & 0.0509 & 0.0885 & 0.833 & 0.0506 & 0.174 \\
    Frozen GFT            & 0.0488 & 0.0953 & 0.853 & 0.0533 & \textbf{0.169} \\
    Frozen GFT (POD)      & 0.0496 & \textbf{0.0879} & \textbf{0.828} & 0.0524 & 0.218 \\
    Learned (gftinit)     & \textbf{0.0463} & 0.0886 & 0.861 & 0.0455 & 0.182 \\
    Learned (fourierinit) & 0.0474 & 0.0894 & 0.852 & \textbf{0.0434} & 0.182 \\
    \midrule
    Ensemble              & 0.0457 & 0.0859 & 0.835 & 0.0448 & 0.182 \\
    \bottomrule
  \end{tabular}
\end{table}
\subsection{Findings}

\paragraph{Frozen physics bases are inconsistent.} The Frozen GFT of
Eq.~\eqref{eq:gamma} is better than the DFT on A ($4\%$,
$p\approx2\times10^{-4}$) but worse on every other large-$n$ stratum
(Table~\ref{tab:main}), and significantly worse under the baseline's
originally released configuration ($p\approx3\times10^{-6}$, three
seeds, paired); Frozen GFT (POD) ties on A ($p=0.64$) and is
directionally worse on C ($p=0.011$).

\paragraph{Learnable bases help in-distribution.}
Both learned models beat the DFT on both in-distribution protocols:
gftinit on 162/200 cases on A ($p\approx5\times10^{-20}$, a $9\%$
reduction) and 135/189 on C-nominal ($p\approx2\times10^{-10}$,
$10\%$); fourierinit on 157/189 on C-nominal
($p\approx3\times10^{-20}$, $14\%$) and 134/200 on A
($p\approx6\times10^{-10}$, $7\%$). B-nominal is a tie.

\paragraph{Under zero-shot shift, learned bases collapse uniformly.}
On B's 41 stressed cases, absent from training, both learned models
lose to the DFT and to Frozen GFT (POD) on \textbf{41/41}
($p\approx9\times10^{-13}$), including stressed cases that remain
stable, so the trigger is regime unfamiliarity, not instability.

\paragraph{Exposure is the cause.} With
stressed cases in training (C), the same architecture beats Frozen
GFT (POD) on 8/11 stressed cases. Zeroing the rotation at inference
recovers 56\% of the stressed gap; span-locked retraining recovers
59\%. Scaling the rotation $t\in[0,1]$ trades nominal for stressed
error monotonically (Fig.~\ref{fig:dial}), and two pre-registered
constrained variants failed to escape the trade
(App.~\ref{app:tech}).

\paragraph{Two trained models give a predictor and an alarm.} The
ensemble of Table~\ref{tab:main} is the best model on A and
B-nominal and, unlike its learned member, does not collapse under
shift (B-stressed $0.835$, near the DFT's $0.833$).
The same pair's seed-averaged per-case disagreement is a label-free
shift detector: it flags all 41 of B's stressed cases with zero
false alarms on the 159 nominal cases (App.~\ref{app:tech}).

\section{Conclusion and outlook}
\label{sec:conclusion}

Making the spatial basis learnable improves accuracy by $7$--$14\%$
over a matched DFT baseline at identical parameters and budget,
across both in-distribution protocols. The gain belongs to
in-distribution operation: under zero-shot regime shift the
adaptation reverses, continuously, and we localized the fragility to
the basis and its co-adapted weights. The resolution requires
nothing new: the ensemble is the best predictor among shift-robust
variants, and seed-averaged model disagreement supplies a label-free
alarm \cite{lakshminarayanan2017} that flagged every shifted case
with no false alarms.

Two directions follow. Scale: our findings are scoped to a modest
budget, and broader training distributions may extend the gain
while shrinking the robustness cost. Topology, unique to the graph
construction: a line outage (the routine N$-$1 contingency) can in
principle be handled by rebuilding $\Gamma$ and swapping in its new
eigenbasis without retraining, a transfer a
topology-blind Fourier basis cannot express. Evaluating this, with
spectral losses targeting missed-instability errors, is our next
step.

\appendix

\section{Technical details}
\label{app:tech}

\begin{figure}[h]
  \centering
  \includegraphics[width=\linewidth]{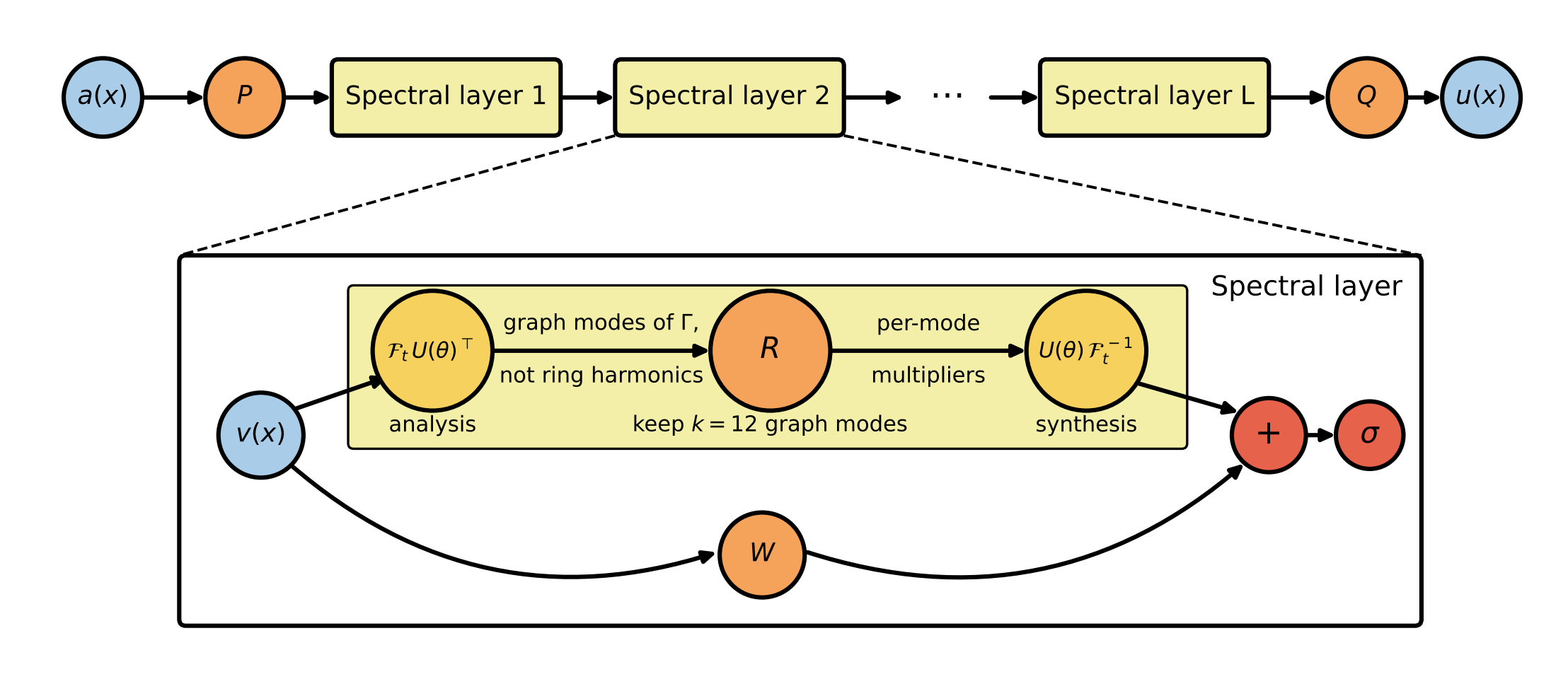}
  \caption{The operator with the generator-axis basis replaced.
  Layout follows the FNO layer diagram of \cite{li2021} and its
  power-system adaptation \cite{cui2024}; the change is in the
  transforms, where the generator-axis DFT is replaced by
  $U(\theta)$, so the retained $k{=}12$ modes are graph modes of
  $\Gamma$ rather than harmonics of the machine ordering. The
  temporal transform $\mathcal{F}_t$ and the bypass path $W$ are
  unchanged.}
  \label{fig:arch}
\end{figure}

\begin{figure}[h]
  \centering
  \includegraphics[width=0.62\linewidth]{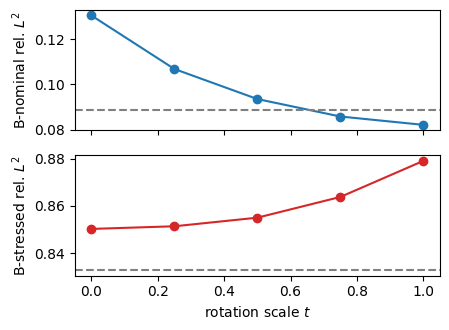}
  \caption{Scaling the learned rotation of the trained gftinit model
  (seed 0). Top: B-nominal mean relative $L^2$ error; bottom:
  B-stressed. Dashed lines mark the DFT reference in each stratum.
  Nominal error improves and stressed error degrades monotonically
  as the rotation is applied.}
  \label{fig:dial}
\end{figure}

\paragraph{Rotation-scaling dial.} Figure~\ref{fig:dial} evaluates
the trained gftinit model with its rotation generator scaled by
$t\in\{0,0.25,0.5,0.75,1\}$, i.e.
$U(t)=\exp\!\big(t\,(W-W^{\top})\big)U_{\mathrm{init}}$, keeping all
other weights fixed (single representative seed, seed 0). B-nominal
error falls and B-stressed error rises monotonically in $t$: the
adaptation is a continuous dial, and no interior point wins both
strata.

\paragraph{Localizing the fragility.} The stressed gap is the
difference in mean B-stressed error between Learned (gftinit) and
Frozen GFT (POD). Two interventions bound the contribution of the
basis itself. Zeroing the rotation at inference while keeping all
co-trained weights recovers $56.3\%$ of this gap (seed 0,
$0.879\to0.850$ against the frozen reference $0.828$). Retraining
with the rotation restricted to the span of the physics basis
recovers $59\%$ (three seeds). Neither intervention closes the gap:
the fragility lies partly in where the basis moved and partly in
weights co-adapted to it, which is consistent with the failure of
both basis-only constrained variants below.

\paragraph{Pre-registered constrained variants.} Two basis-only
restrictions were trained against gates fixed in advance: V0 locks
the rotation to the span of the physics basis (55 degrees of
freedom); V1 permits rotation only within clusters of dynamically
similar modes (465). V0 fails both gates (B-stressed $+0.014$
vs.\ frozen, $p\approx5\times10^{-9}$; B-nominal $+0.003$ vs.\ the
unconstrained model). V1 keeps the nominal gain but fails
robustness ($+0.022$, $p\approx2\times10^{-12}$). Both overall
failures follow from the split localization above: no basis-only
constraint can repair fragility that lives partly in the co-adapted
weights.

\paragraph{Disagreement alarm.} Per-case disagreement between
Learned (gftinit) and Frozen GFT (POD) is the relative $L^2$
distance between their predictions,
$d_i = \|\hat{y}_i^{\mathrm{L}}-\hat{y}_i^{\mathrm{F}}\|_2 \,/\,
\tfrac{1}{2}(\|\hat{y}_i^{\mathrm{L}}\|_2+\|\hat{y}_i^{\mathrm{F}}\|_2)$,
averaged over the three seed pairings. On Protocol~B the resulting
nominal and stressed distributions are disjoint: the maximum over
the 159 nominal cases is $0.206$ and the minimum over the 41
stressed cases is $0.327$, so any threshold between them flags all
41 shifted cases with zero false alarms. Seed averaging matters:
individually, seeds 0 and 2 separate the strata completely while
seed 1 narrowly overlaps (per-seed AUROC $0.9998\pm0.0003$), so the
perfect alarm is a property of the seed-ensembled statistic rather
than of any single pair.

\end{document}